\documentclass[square]{ws-procs11x85}
\usepackage{balance} 
\usepackage{pst-grad}
\def\Journal#1#2#3#4{{#1} {\bf #2}, #3 (#4)}

\newcommand{\met}{\hbox{E\kern-0.5em\lower-0.1ex\hbox{/}}_T}
\newcommand{\diff}{\textrm{d}}

\begin{document}

\twocolumn[
\title{Environmental limits on the non-resonant cosmic-ray 
current-driven instability}

\author{B. Reville, J.G. Kirk}

\address{Max-Planck-Institut f\"ur Kernphysik, Heidelberg 69029}
\author{P. Duffy}

\address{UCD School of Physics, University College Dublin, Belfield, Dublin 4}

\author{S. O'Sullivan}
\address{School of Mathematical Sciences,
Dublin City University,
Glasnevin,
Dublin 9}

\begin{abstract}
We investigate the so-called non-resonant cosmic-ray streaming
instability, first discussed by Bell (2004). The extent to which
thermal damping and ion-neutral collisions reduce the growth of this
instability is calculated. 
Limits on the growth of the non-resonant mode in SN1006 and
RX J1713.7-3946 are presented.
\end{abstract}
\vskip12pt  
]

\bodymatter

\section{Introduction}

Diffusive shock acceleration at the outer shocks of supernova remnants
is believed to be the primary source for galactic cosmic rays.  
In the simplest picture, however, the maximum energy attainable
by this mechanism falls short of the so-called knee at approx.
$1$ PeV\cite{LagageCesarsky}. Amplification of the magnetic field 
in the vicinity of shock front may facilitate acceleration to 
higher energies, without changing the simplicity of the model.
There has also been an increase in the wealth of observational 
evidence for magnetic field amplification at collisionless shocks,
such as the non-thermal bright X-ray rims in several 
supernova remnants\cite{VinkLaming, voelk}. The growth of the
magnetic field is thought to be closely related to the
efficient production of cosmic rays and the
non-resonant current-driven instability\cite{Bell2004}
has provided a mechanism which may indeed considerably
amplify the magnetic field. 

Generally speaking, the medium upstream of a supernova blastwave
will not be completely ionised.
The presence of a small neutral component limits the
acceleration of cosmic rays in these environments\cite{DruryDuffyKirk}, 
due to damping of resonantly excited Alfv\'en waves.
In molecular clouds, where the ionisation
fractions are even lower and the densities much higher, the situation
is even more extreme\cite{KulsrudPearse,ZweibelShull}. In a certain 
range of wavelengths, Alfv\'en waves simply do not propagate, and are quite
rapidly damped.

A non-resonant mode, driven by a cosmic-ray current in a 
partially ionised medium has recently been studied\cite{Bykov}.
Treating the background particles as a MHD single fluid 
with a generalised Ohm's law, and using a kinetic description of 
the cosmic rays, these authors conclude that a non-resonant mode
with a rapid growth rate remains. However, 
as we show, for high frequency waves
the friction between the charged species and the neutrals is
not sufficient to couple their motions, and a single fluid
description of the plasma is no longer appropriate.

In the present work, we investigate various conditions that may
reduce the growth of the non-resonant instability. We first calculate
the effect of thermal damping of the waves. We also determine an 
expression for the dispersion relation of waves in the presence
of an arbitrary fraction of neutral particles. These results 
are then used to speculate on whether or not magnetic field amplification
may have taken place in SN1006 or in the molecular cloud close to
RX J1713.7-3946. We conclude with some remarks about the importance of
the non-resonant current-driven instability.


\section{Thermal effects}

We present here a simple derivation of the cosmic-ray current-driven 
instability. Since the fastest growing mode is non-resonant we do not
need to calculate the full effects of an anisotropic cosmic-ray
distribution\cite{reville07}, 
and we investigate here a super-Alf\'enic beam, which
considerably simplifies the analysis.

The linear dispersion relation for circularly polarised transverse 
waves propagating parallel to the zeroth order magnetic field is
\begin{equation}
\label{disprel}
\frac{c^2k^2}{\omega^2}-1 = \sum_s \chi_s(k,\omega) ,
\end{equation}
where the summation is over each species with corresponding 
charge $q_s$, 
cyclotron frequency $\omega_{\rm cs} = q_{s}B_0/m_{s}c$
and plasma frequency $\omega^2_{\rm ps} = 4\pi q_s^2 n_{s}/m_{s}$.
The susceptibility $\chi_s$ for each 
component of a plasma is determined by integrating
the Vlasov equation along the unperturbed trajectories
about the zeroth order field and can be found in any standard textbook
on plasma physics (e.g. Krall \& Trivelpiece 1973).
For circularly polarised waves propagating parallel to a uniform magnetic
field, the susceptibility for each species is given by
\begin{eqnarray}
\label{yoon_sus}
\chi_s = \frac{4 \pi q_s^2}{\omega^2}\int \diff^3 p
\frac{v_\bot p_\bot}{\omega \pm \omega_{\rm cs}-k v_\parallel}
\;\;\;\;\;\;\;\;\nonumber\\
\left[ (\omega - k v_\parallel)
\frac{\partial f_{s} }{ \partial p_\bot^2}+
 k v_\parallel 
\frac{\partial f_{s} }{ \partial p_\parallel^2}\right].
\end{eqnarray}

For simplicity, the plasma we consider has only three components.
A two species thermal population of electrons and protons, and a cosmic
ray component which we assume to consist entirely of protons. Since 
we are investigating waves on lengthscales longer than the Debye length
the plasma can be taken to be quasi-neutral, resulting in a 
slight charge excess
of electrons in the thermal plasma due to the presence of
cosmic rays. A neutralising 
background flux is also induced to balance the current produced
by the streaming cosmic rays.

Assuming a Maxwellian background and a mono-energetic beam of 
protons with momentum $p_0$ 
directed along the magnetic field, it is straightforward to show 
that the dispersion relation for low frequency waves 
$|{\omega}| \ll \omega_{\rm ci} < |\omega_{ce}|$
is
\begin{eqnarray}
\label{disp}
\frac{\omega^2}{ k^2c^2}+
\epsilon \frac{{\omega}}{\omega_{\rm ci}}
\Theta-\frac{\rm v_A^2}{c^2}-
 \frac{\zeta {\beta}_{\rm b}^2}{1-\epsilon kr_{\rm g}}
=0~
\end{eqnarray}
where $c\beta_{\rm b}$ is the speed of the proton beam, 
${\rm v_A} = B_0/\sqrt{4\pi n_{\rm i} m_{\rm i}}$ is the Alfv\'en velocity
and $\Theta = k_B T_{\rm i}/m_{\rm i} c^2$ the dimensionless ion temperature. 
We have also introduced the parameter $\epsilon$ to describe
the polarisation with $\epsilon=+1(-1)$ for right (left)-handed waves ($\omega>0$).

We define $r_{\rm g} = p_{0}c/eB_0$ the gyroradius of the 
protons in the beam and consider only waves with 
$k>0$. 
We have also 
introduced a dimensionless parameter characterising
the strength of the driving term
\begin{equation}
\zeta = 
\frac{n_{\rm cr} p_{0}}{n_{\rm i} m_{\rm i} {\beta}_{\rm b} c}
\nonumber
\end{equation}

The maximum growth rate of the non-resonant mode is
\begin{equation}
\label{gmax}
 {\rm Im}(\omega)
\approx\frac{1}{2}\frac{c}{{\rm v_A}}
\frac{n_{\rm cr}}{n_{\rm i}}\beta_{\rm b}\omega_{\rm ci},
\end{equation}
which is independent of magnetic field strength.

We can now determine possible thermal effects on the growth rate.
For $k r_{\rm g} \gg 1$, it follows from (\ref{disp}),
in the limit $\Theta \gg {\rm v_A}^2/c^2$, that the maximum growth rate is
\begin{equation}
{\rm Im}(\omega) \sim \left({n_{\rm cr}}/{n_{\rm i}}\right)^{2/3}
\left({\beta_{\rm b}^2}/{\Theta}\right)^{1/3}\omega_{\rm ci}.
\end{equation}
 
In order for the non-resonant mode to leave the 
regime of linear growth before being overtaken by the shock front, i.e., 
before being advected over a distance of roughly $r_{\rm g}/{\beta}_{\rm b}$,
one requires 
${\rm Im}(\omega) > {\beta}_{\rm b}^2 c/r_{\rm g} $.
The necessary condition for thermal effects to reduce the growth 
rate below this value is
\begin{equation}
 \Theta >  \frac{\zeta^2 p_{0}}{\beta_{\rm b}^{2}m_{\rm i} c }.
\end{equation}
For typical SNR parameters this condition will only be
satisfied for a very weak driving term , $\zeta \ll 1$. 
However, thermal effects are likely to play a significant role
for relativistic shocks, and may even provide a saturation mechanism
for the current driven instability \cite{revilleetal}.


\section{Collisional effects}

The surroundings into which supernova shocks propagate are,
in general, not completely ionised.
Observations have shown that many supernova take place near 
molecular clouds, which can have very low ionisation fraction.
We consider the reduction of the 
growth of the non-resonant modes due to ion-neutral friction.
For wave frequencies larger than the momentum exchange 
frequency, a multi-fluid treatment of the plasma is necessary,
since the collisions are not frequent enough to couple the
motion of the neutral particles to the charged ions. 
We consider two fluids, a single MHD fluid for the charged species
and a neutral component.   
A simple analysis of the frictionally 
coupled system with an external cosmic ray current
results in the following dispersion relation\cite{reville07}
\begin{equation}
\label{dispIN}
 \omega^2\left(1+\frac{i\nu_{\rm in}}{\omega+i\nu_{\rm ni}}\right)
= k^2{\rm v_A}^2 + \epsilon\zeta\frac{{\rm v}_{\rm s}^2}{r_{\rm g}}k(\sigma-1).
\end{equation}
where $\nu_{ab}$ is the momentum exchange frequency from species $a$ to
species $b$ and are related by momentum conservation 
$\rho_{\rm n}\nu_{\rm ni} = \rho_{\rm i}\nu_{\rm in}$.
The complex function $\sigma(k)$ represents the susceptibility of the 
cosmic rays normalised to the induced thermal return
current\cite{Bell2004,revilleetal}.
For plasmas in the temperature range $10^2~{\rm K} < T < 10^5~{\rm K}$,
\cite{KulsrudCesarsky} give the following expression
\begin{equation}
\label{nuin}
 \nu_{\rm in} \approx8.9\times 10^{-9} n_{\rm n}
\left(\frac{T}{10^4~{\rm K}}\right)^{0.4} ~{\rm s}^{-1}.
\end{equation}

In the high frequency limit, $|\omega| \gg \nu_{\rm in}$, this reduces
to a dispersion relation similar to Eq. (\ref{disp}). 
In the low frequency limit 
$|\omega| \ll \nu_{\rm ni}$ we find
\begin{equation}
\label{denseDR}
 \omega^2 \approx\frac{\rho_{\rm i}}{\rho} 
\left[k^2{\rm v_A}^2 + \epsilon\zeta\frac{{\rm v}_{\rm s}^2}{r_{\rm g}}k(\sigma-1)\right].
\end{equation}
If the collision frequencies are very large, 
the neutral and ionised components are tied together and the 
effect of ion-neutral collisions is simply 
to increase the effective mass of the ions. 

For strongly driven, non-resonant modes Eq. (\ref{dispIN}) reduces to 
\begin{equation}
 \omega(\omega^2+\zeta\frac{{\rm v}_{\rm s}^2}{r_{\rm g}}k)
+i\nu_{\rm in}\left[(1+Z)\omega^2+Z\zeta\frac{{\rm v}_{\rm s}^2}{r_{\rm g}}k
\right]=0,
\end{equation}
where $Z=\rho_{\rm i}/\rho_{\rm n} =\nu_{\rm ni}/\nu_{\rm in}$. For
all physically relevant parameters, this cubic
polynomial in $\omega$ has three purely imaginary roots, two of which
are damped modes, and the third the non-resonant growing mode.  Thus,
ion-neutral 
collisions are unable to stabilise the strongly driven mode, although
they can affect its growth rate. Interestingly this result also applies
to any aperiodic mode.

The general analytic expressions for the growing mode are cumbersome, but
various limiting cases yield interesting results.  
In the limit of low ionisation, $Z\gg1$ one recovers
the growth rate given in Eq. (\ref{gmax}). 
For $Z\ll1$ the 
growth rate is given by
\begin{equation}
 \gamma=-\frac{\nu_{\rm in}}{2}+
\frac{1}{2}
\sqrt{\nu_{\rm in}^2+4\zeta\frac{{\rm v}_{\rm s}^2}{r_{\rm g}}k}
\end{equation}
If $\nu_{\rm in}^2\ll4\zeta{\rm v}_{\rm s}^2k/r_{\rm g}$ collisions are too slow
to compete with the driving, and  
the growth rate is again
the same as Eq. (\ref{gmax}) with a negligible reduction due to collisions.
However, in addition to the low frequency limit in which the components are
tied together Eq. (\ref{denseDR}), a new regime arises. If the collisional
drag on the ions is sufficient to affect the driving, but not to couple the components, i.e., 
$\nu_{\rm in}^2\gg 4\zeta{\rm v}_{\rm s}^2k/r_{\rm g}\gg \nu^2_{\rm ni}$, 
then we obtain 
\begin{equation}
 \gamma\approx \zeta\frac{{\rm v_s}^2}{\nu_{\rm in} r_{\rm g}}k=
 \frac{\omega_{\rm ci}}{\nu_{\rm in}}\frac{n_{\rm cr}}{n_{\rm i}}k{\rm v_s} .
\end{equation}
A similar result to this has been found previously\cite{Bykov}, but as
we have shown here, this is valid only when $\gamma\ll\nu_{\rm in}$.
\begin{figure}
\begin{center}
  \includegraphics[width=0.9\columnwidth]{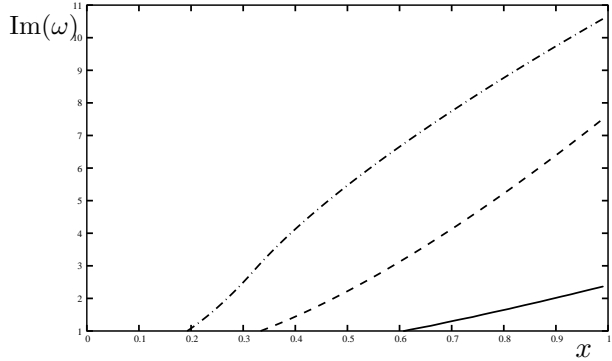}
\rput(-0.5,-0.1){\large{$x$}}
\rput(-7.67,4.2){Im($\omega$)}
\caption{%
Maximum growth rate as a function
of ionisation fraction for total density $n=n_{\rm i}+n_n$ 
in a H-H$^+$ gas: $0.1{\rm~cm}^{-3}$, $10^4$\,K  (solid),
$1.0{\rm~cm}^{-3}$, $10^3$\,K (dash), $10.0{\rm~cm}^{-3}$, $10^2$\,K (dash-dot).
We take $\zeta=0.01x_{\rm i}{\rm v}_{\rm s}/c$ with
shock speed ${\rm v_s}=5000{\rm ~km~s}^{-1}$, and $r_{\rm g}=10^{14}{\rm cm}$.
$\omega$ is in units of ${\rm v}_{\rm s}^2/r_{\rm g}c$. Once the 
$\omega_{\rm max} \leq 1$ the non-resonant instability is not
effective in amplifying the magnetic field. 
}
\label{fig_x}
\end{center}
\end{figure}

In Fig.~\ref{fig_x} we illustrate the influence of 
ion-neutral collisions for SNR parameters using a full numerical
solution to the dispersion relation. Thermal damping
is negligible, but a weak dependence on temperature enters via
Eq. (\ref{nuin}). We plot the maximum growth
rate, as a function of ionisation fraction $x_{\rm i}\equiv n_{\rm i}/n$,
taking the driving term $\zeta=0.01 x_{\rm i} {\rm v_s} /c$
since the cosmic rays couple directly only to the ionised component.
For higher density plasmas, a larger fraction of neutral 
particles is necessary to reduce the growth rate below the threshold
value of ${\rm v}_{\rm s}^2/r_{\rm g}c$.
 

\section{Application to SNR}

We now investigate the role of the non-resonant instability in some
typical supernova remants. Although it is beyond the scope of this work to
estimate the possible saturated field strength from the non-resonant 
instability, we can nevertheless test the consistency of the analysis, by
checking that the instability can enter the non-linear regime before 
being overtaken by the shock.
Thus, the physical conditions far upstream ($B$, $n$, $x_{\rm i}$, $T$) together with the shock speed
${\rm v}_{\rm s}$ and cosmic ray intensity $U_{\rm cr}/\rho_{\rm i}{\rm v_s}^2=0.1$,
should combine to yield a growth rate in excess of ${\rm v_s}^2/(r_{\rm g}c)$.
We investigate two extreme cases.

Firstly, for SN1006, the shock velocity is approximately ${\rm v_s} \approx2900{\rm~km~s}^{-1}$
the density is $0.05 \lesssim n \lesssim 0.25 {\rm ~ cm}^{-3}$, and the neutral fraction
is small, $x_{\rm i}\approx0.9$ \cite{Raymond}. Taking 
$n_{\rm i}=0.25 {\rm ~ cm}^{-3}$ gives 
$\zeta{\rm v_s}^2/{\rm v_A}^2 \sim 10$.  The low density and shock speed, 
imply that
the non-resonant mode is not very strongly driven,  
but the driving is nevertheless strong compared to the collision frequency.  
Very close to the shock, the growth time of the field is on the order of
years, while further from the shock 
the growth rate decreases quite dramatically, as
the number density of cosmic rays decreases.
For the parameters given above the maximum growth rate is 
$\sim 10 {\rm v_s}^2/r_{\rm g} c$. The non-resonant mode may still amplify 
the field into the nonlinear regime.

The low neutral fraction in SN1006 means that ion-neutral collisions
are not likely to
dramatically reduce the instability. The interaction
of a supernova blast wave with a molecular cloud, 
such as that observed close to
the north-western rim of RX~J1713.7-3946
would have a much 
more significant effect on the growth. 
Although the exact parameters are uncertain,
molecular clouds are generally clumpy with interclump densities in the range 
$5-25 {\rm~cm}^{-3}$ and ionisation fractions not larger than $10\%$
\cite{Chevalier99}. Adopting the values in \cite{Malkovetal}
$T=10^2{\rm~K},~n=23{\rm~cm}^{-3}, ~x_{\rm i}=0.01, {\rm~v_s}=10^8{\rm cm~s}^{-1}$
the collision frequency is larger than the growth rate.
The non-resonant driving term is unable to dominate
over the Alfv\'en term and growth is driven by the
cosmic rays themselves, at very long wavelengths $kr_{\rm g}\ll1$.
At shorter length scales, the waves are rapidly ion-neutral damped Alfv\'en
waves. The shortest growth timescale is on the order of $\sim 10^6$ yrs. 
This value varies with choice of $r_{\rm g}$, temperature and density,
but for high density and low ionisation fractions, the growth 
timescale is longer than the free expansion phase of a 
supernova remnant, or the lifetime of the wave in the shock precursor,
suggesting that the Fermi acceleration mechanism 
may switch off upon interaction with a molecular cloud.


\section{Discussion}

It is now widely believed that magnetic field amplification is
the best method for pushing the maximum energy beyond the 
Lagage--Cesarsky limit\cite{LagageCesarsky}. The non-resonant 
instability has been the most promising mechanism so far 
in that its growth rate is considerably faster than its
resonant counterpart, and initial studies suggest that it does not 
saturate at a level $\delta B \lesssim B_0$. However, since the
mechanism is non-resonant, with the fastest growing 
modes at very short wavelengths, 
the amplified field has very little effect on the acceleration of
the maximum energy particles unless the energy is transferred to
the large scale field by some non-linear process,
as seen in some numerical simulations\cite{Bell2004, Riquelme}.
The results reported on in this paper extend the previous 
studies and allow application to more extreme environments.
Such processes may be crucial for understanding 
particle acceleration at supernova remnant shocks in the
galactic centre region, for example. 
Efficient acceleration is essential for this
mechanism to occur. The largest uncertainty in our model
arises in determining the value of the driving term $\zeta$. 
It is suggested, however, that efficient acceleration naturally
leads to magnetic field amplification.

\section*{Acknowledgments}
This research was jointly supported by 
COSMOGRID and the Max-Planck-Institut f\"ur Kernphysik, Heidelberg.
The authors would like to thank the conference organisers.

\end{document}